\documentclass[preprint,12pt]{elsarticle}

\usepackage{multicol}
\usepackage{graphicx}
\usepackage{booktabs}
\usepackage{amssymb,bm,mathrsfs,bbm,amscd}
\usepackage[tbtags]{amsmath}
\usepackage{lastpage}
\usepackage{CJK}
\usepackage{color}
\usepackage{enumerate}

\usepackage{url}

\usepackage[flushleft]{caption2}
\usepackage{epic,eepic}

\usepackage{keyval,graphicx}
\usepackage{overpic}
\usepackage{tikz}
\usetikzlibrary{arrows,snakes,backgrounds}
\usetikzlibrary{patterns}
\usepgflibrary{shapes.arrows}

\usepgflibrary{decorations.pathmorphing}
\usepgflibrary{decorations.pathreplacing}
\usepgflibrary{decorations.shapes}
\usepgflibrary{decorations.fractals}

\usepackage{scalefnt}
\usepackage{lscape}
\usepackage{wrapfig}
\usepackage{picinpar}

\usepackage{cases}
\usepackage{pifont}

\usepackage{times}
\usepackage{scrtime}

\usepackage{lineno}

\usepackage{array}
\newcommand{\PreserveBackslash}[1]{\let\temp=\\#1\let\\=\temp}
\newcolumntype{C}[1]{>{\PreserveBackslash\centering}p{#1}}
\newcolumntype{R}[1]{>{\PreserveBackslash\raggedleft}p{#1}}
\newcolumntype{L}[1]{>{\PreserveBackslash\raggedright}p{#1}}

\usepackage{xcolor}

\newcommand{\wuhao}{\fontsize{10.5pt}{\baselineskip}\selectfont}

\newcommand{\liuhao}{\fontsize{7.875pt}{\baselineskip}\selectfont}

\newcommand{\cpp}{C\raisebox{1.5pt}{\scalebox{0.7}{++}}}

\newcommand{\ie}{\emph{i}.\emph{e}.}

\newcommand\figcaption{\def\@captype{figure}\caption}
\newcommand\tabcaption{\def\@captype{table}\caption}
\renewcommand\figurename{Fig.}

\usepackage{amssymb}

\newcounter{bla}

\definecolor{doushalu}{RGB}{199,238,206}
\definecolor{darkblue}{RGB}{0,112,192}

\journal{arXiv.org}

\begin{document}


\begin{frontmatter}

\title{NeuDATool: An Open Source Neutron Data Analysis Tools, Supporting GPU
Hardware Acceleration, and Across-computer Cluster Nodes Parallel}

\author[add1,add2]{Changli Ma}
\author[add1,add2]{He Cheng\corref{author}}
\author[add1,add2]{Taisen Zuo}
\author[add1,add2]{Guisheng Jiao}
\author[add1,add2,add3]{Zehua Han}
\cortext[author] {Corresponding author.\\\textit{E-mail address:}
machangli@ihep.ac.cn (Changli Ma), chenghe@ihep.ac.cn (He Cheng), zuots@ihep.ac.cn (Taisen Zuo),
 jiaogs@ihep.ac.cn (Guisheng Jiao), hanzh@ihep.ac.cn (Zehua Han)}
\address[add1]{Institute of High Energy Physics, Chinese Academy of Sciences (CAS)，Beijing 100049, China}
\address[add2]{Spallation Neutron Source Science Center, Dongguan 523803, China}
\address[add3]{University of Chinese Academy of Sciences, Beijing 100049, China}

\begin{abstract}

Empirical potential structure refinement (EPSR) is a neutron scattering data
analysis algorithm and a software package. It was developed by the British
spallation neutron source (ISIS) Disordered Materials Group in 1980s, and aims to
construct the most-probable atomic structures of disordered liquids.
It has been extensively used during the past decades, and has generated
reliable results. However, it is programmed in Fortran and
implements a shared-memory architecture with OpenMP.
With the extensive construction of supercomputer clusters and the widespread
use of graphics processing unit (GPU) acceleration technology, it is now necessary
to rebuild the EPSR with these techniques
in the effort to improve its calculation speed. In this study, an open source framework
NeuDATool is proposed. It is programmed in the object-oriented language C++,
can be paralleled across nodes within a computer cluster, and supports GPU acceleration.
The performance of NeuDATool
has been tested with water and amorphous silica neutron scattering data. The test shows
that the software could reconstruct the correct microstructure of the samples, and the
calculation speed with GPU acceleration could increase by more than 400 times compared
with CPU serial algorithm at a simulation box consists about 100 thousand atoms.
NeuDATool provides another choice
for scientists who are familiar with C++ programming and want to
define specific models and algorithms for their analyses.

\end{abstract}


\end{frontmatter}


\section{Introduction}
\label{Introduction}
Neutron total scattering for disordered materials is a powerful tool to study the
most probable atomic structures in an amorphous system. Since the introduction of the first
total scattering spectrometer (TSS)\cite{TSS_01}, which was developed at HELIOS in the 1970s, numerous
important scientific problems have been solved with the use of these types of instruments,
including the elucidation of the structures of water at high- and low-densities, and the
observation of the heterogeneities in mixed alcohol--aqueous solutions.

The success of neutron total scattering for disordered materials is based on deuteration
techniques. The scattering pattern $S(Q)$ is a weighted summation of the Fourier
transforms of all the pair correlation functions (PDF),
 {\wuhao
\begin{equation}
\label{eq:cs_cal_liq}
S(Q) = \sum_{\alpha=1}c_{\alpha}b_{\alpha}^{2} + \sum_{\alpha= 1,\beta\geq\alpha}
(2-\delta_{\alpha\beta})c_{\alpha}c_{\beta}b_{\alpha}b_{\beta}\{4\pi\rho
\int_{0}^{\infty}r^{2}(\textsf{g}_{\alpha\beta}(r)-1)\frac{\sin(Qr)}{Qr}dr \}
\end{equation}
}
where $Q$ is the scattering vector, $\sum_{\alpha}c_{\alpha}b_{\alpha}^{2}$ is a flat background,
and $(2-\delta_{\alpha\beta})c_{\alpha}c_{\beta}b_{\alpha}b_{\beta}$ is the
weighted factors of different partial structural factors,
$c_{\alpha /\beta}$ and $b_{\alpha/\beta}$ are the atom ratio and scattering
length of atoms with types $\alpha/\beta$, respectively.
$g_{\alpha\beta}(r)$ is the PDF between atom types $\alpha$ and $\beta$.
If the different atom type number is $M$ in a sample,
there are $M(M+1)/2$ different $g_{\alpha\beta}(r)$ functions in
\textbf{Eq. (\ref{eq:cs_cal_liq})}, and we must solve all of them first before
the most probable atomic structure is obtained.
It is thus difficult to reveal the most probable all-atom structure of disordered
material, just according to one scattering curve.
Fortunately, neutron scattering scientist has deuteration technique.  Because deuterated
samples have almost the same atomic structure as their hydrogenate forms, each
deuterated sample can thus generate a different scattering pattern.

All-atom model simulations, such as EPSR and RMC, are common methods to solve the matrix of
\textbf{Eq.(\ref{eq:cs_cal_liq})}s, and reconstruct the atomic structure of the tested
samples\cite{EPSR_01,RMC_03,RMC_04}. Because EPSR can provide a reliable and visualized
atomic microstructure, it has been extensively used.
However, EPSR still contains constraints which limit its potential applications. First it is
programmed in Fortran. Fortran can implement the execution of algorithms with very fast
calculation speeds, but it is a procedure-oriented language. Second, EPSR is paralleled
in shared memory architecture with OpenMP\cite{SMA_01, OpenMP_01}. It cannot be
paralleled across different nodes of a supercomputer cluster. This restricts the calculation
speed and the analysis system scale, such as in the case of a macromolecular system.
A macromolecule generally contains hundreds of atoms. To cover the entire range of
scattering vectors that characterize the entire set of conformation states of macromolecules
in a typical total scattering instrument, the simulation box composed of more than half a
million atoms should be larger than 10nm. EPSR cannot run in such a large
system. Therefore, parallel calculations are necessary.

In light of this, the object-oriented language \cpp { } and the compute unified device
architecture (CUDA)\cite{CUDA_01, CUDA_02}
are used to develop a toolkit NeuDATool.
In NeuDATool, users can define easily a new simulation box, atoms, molecules, and movement
models using the class multiple inheritance mechanism. Graphics processing unit (GPU)
hardware acceleration\cite{GPU_01,GPU_02,GPU_03,GPU_04} is
supported by CUDA C, and this allows the program to take advantage of commonly used GPU
computing servers. In addition, with the distributed memory architecture
API message passing interface (MPI) mpich2\cite{SMA_01,MPI_01}, NeuDATool can be paralleled
across nodes of a supercomputer cluster. These acceleration techniques make the program have
huge speed promotion.

\section{NeuDATool simulation method and program Flow}

\subsection{Algorithmic Principle}

\begin{figure}
\begin{center}
\includegraphics[height=0.5\linewidth,width=0.85\linewidth]
{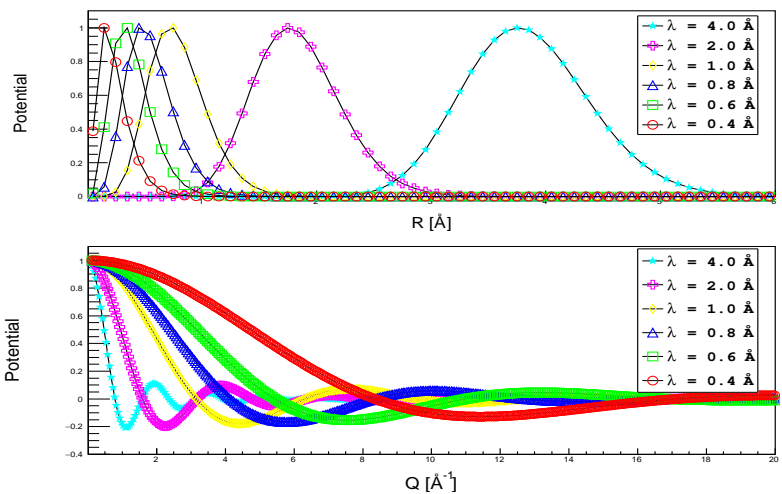}
\renewcommand{\figurename}{Fig.}
\textcolor[rgb]{0,0,1.0} {\figcaption
{
  \label{fig:figure_pot_poisson} A series of Poisson distributions with different $\lambda$
  values in R space (Top) and Q space (Bottom). The distribution list in Q space is used to fit $\Delta S(Q)$.
  The peaks of these distributions are normalized to 1.0.
}
}
\end{center}
\end{figure}

\begin{figure}
\begin{center}
\includegraphics[height=0.5\linewidth,width=0.49\linewidth] {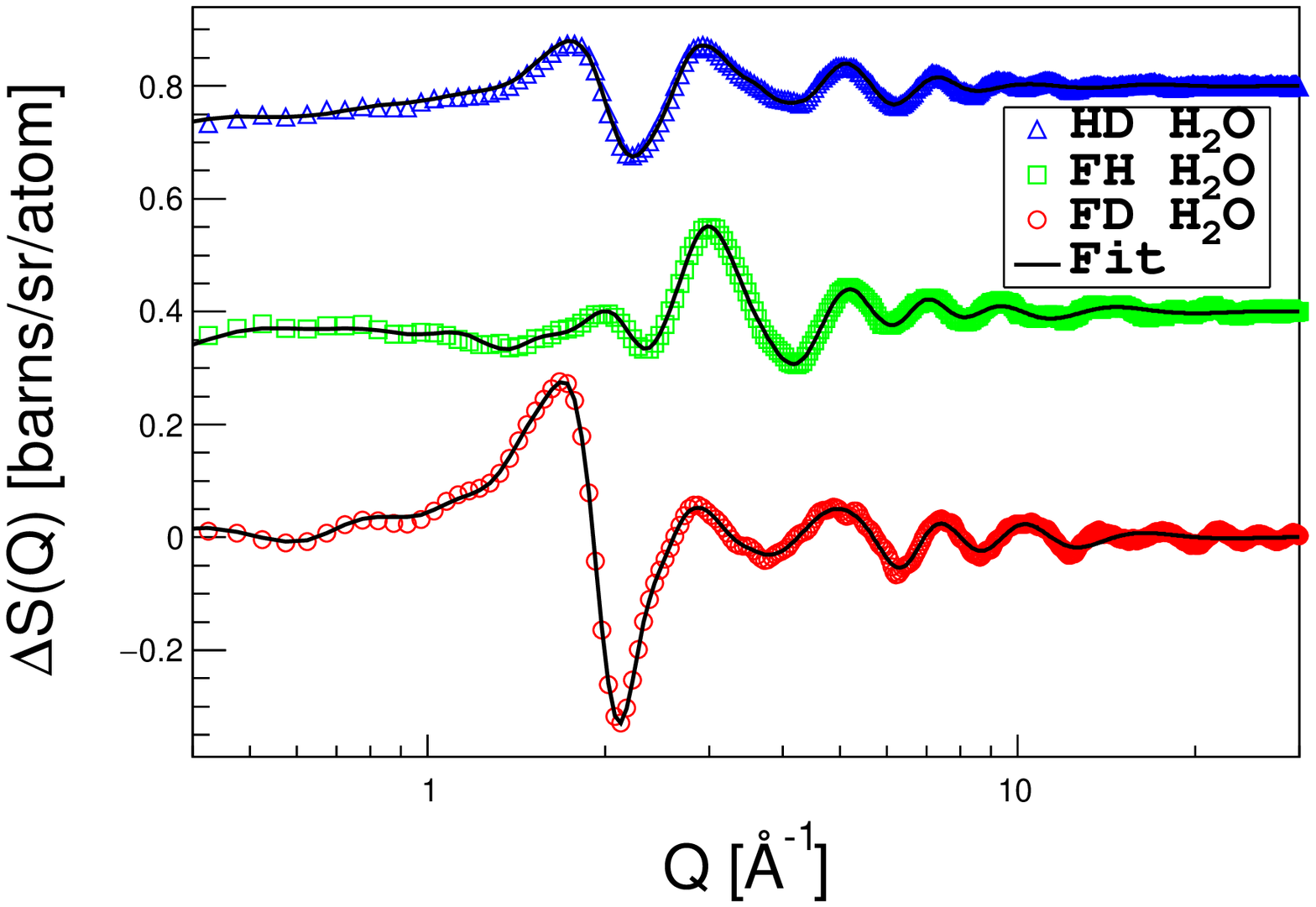}
\includegraphics[height=0.5\linewidth,width=0.49\linewidth] {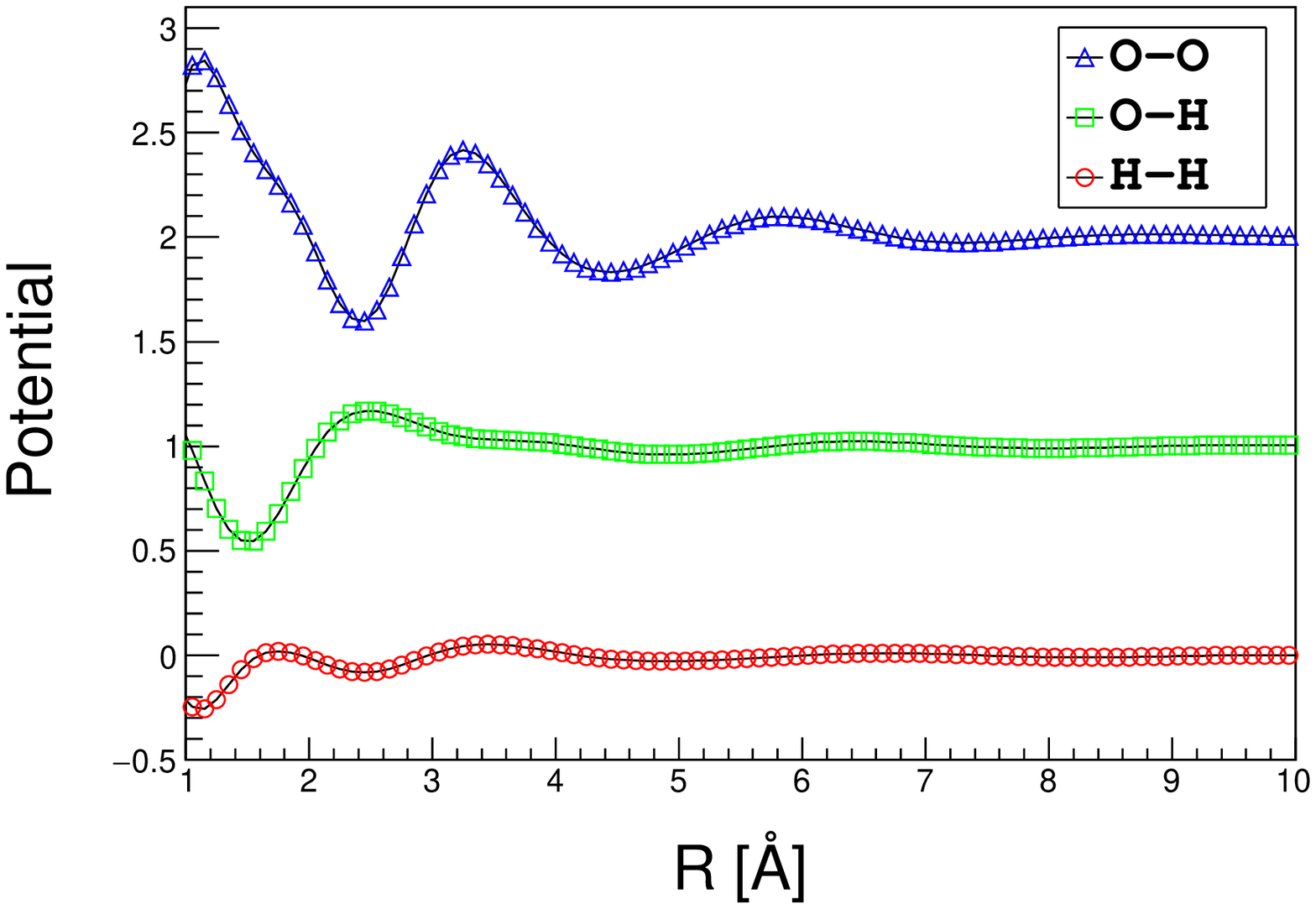}
\renewcommand{\figurename}{Fig.}
\textcolor[rgb]{0,0,1.0} {\figcaption
{
  \label{fig:figure_pot_h2o_emp} The $\Delta S(Q)$ fitting results of H$_2$O samples
  (HD H$_2$O means half deuterated HDO, FH H$_2$O is fully hydrogenated H$_2$O and FD H$_2$O
  is fully deuterated D$_2$O.) with Poisson distribution (left) and the corresponding
  empirical potentials (right). The $\Delta S(Q)$ of FH and HD are shifted by 0.4
and 0.8, respectively. The EMP of O-H and O-O are shifted by 1.0 and 2.0, respectively.
}
}
\end{center}
\end{figure}

NeuDATool is essentially a Monte-Carlo simulation method, but its difference from metropolis MC is
based on the fact that the atomic potentials used include neutron
scattering data information\cite{EPSR_01,EPSR_02}. In NeuDATool, the atomic potential used in
the MC simulation is divided into two categories \ie, ``reference potential (RP)'' and
``empirical potential (EMP)''\cite{EPSR_01}. RP is similar to that used in molecular
dynamics (MD) simulations, so the potential form and parameters can be obtained from
all-atom MD force fields, such as the optimized potential for liquid simulations
(OPLS)\cite{OPLS_01,OPLS_02}, assisted model building with energy refinement (AMBER)\cite{AMBAR_01},
chemistry at HARvard molecular mechanics (CHARMM)\cite{CHARMM_01}, and
others\cite{EPSR_01,Force_Field_01}. By contrast, EMP has no fixed form and it is used to reflect
the differences of neutron structural factors between experiments and MC simulations ($\Delta S(Q)$).
To be exact, EMP is the reverse Fourier transform of $\Delta S(Q)$ as the perturbation to the RP to
guide the simulation approach to scattering measurements. In the present form of NeuDATool,
EMP is expressed based on a list of Poisson distributions in real space and their corresponding
Fourier transforms in Q space\cite{EPSR_03}. Fig.\ref{fig:figure_pot_poisson} shows
a series Poisson distributions with different $\lambda$ values in R space and Q space (top and bottom)
respectively. Here, $\lambda$ is the mathematical expectation of the distribution.
In NeuDATool, the Poisson series in Q space are used to fit $\Delta S(Q)$. Fig.\ref{fig:figure_pot_h2o_emp}
gives examples of the $\Delta S(Q)$ fitting results of H$_2$O samples' simulation (left) and its
corresponding EMP potential (right) respectively.  In short, RP is used to assign molecules with reasonable
shapes and realize other constraints whose correctness have been approved, while EMP is used as a
feedback parameter to lead the MC simulation in a consistently
progressive manner with experimental data. In the simulations, only RP is used in
the MC simulation at the beginning, and the potential changes of the system ($\Delta U$)
are used as the selection criteria of molecules' or atoms' random movements. When the simulation
reaches equilibrium, EMP is introduced to fit $\Delta S(Q)$ and is added to RP to continue
the simulation. When the MC simulation with updated potential reaches equilibrium again,
NeuDATool calculates EMP and updates the simulation potential once more. This process is repeated
until the EMP becomes equal to zero, {\it i.e.}, $\Delta S(Q)$ becomes very small.

\subsection{NeuDATool Algorithmic Flow}

The algorithmic flow of NeuDATool is shown in Fig.\ref{fig:figure_process_flow}.
Details of some processes are described below:

\begin{figure}
\begin{center}
\includegraphics[height=1.1\linewidth,width=0.85\linewidth]
{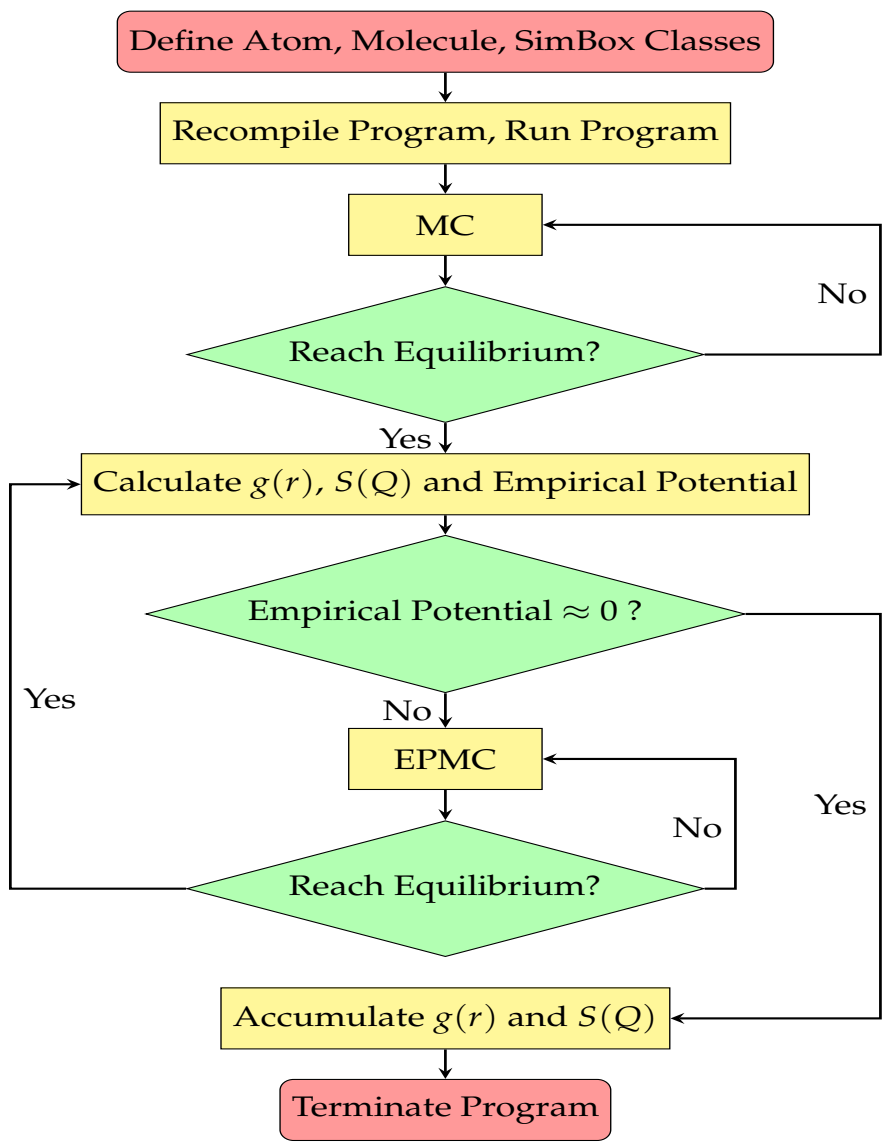}
\renewcommand{\figurename}{Fig.}
\textcolor[rgb]{0,0,1.0} {\figcaption
{
  \label{fig:figure_process_flow} Schematic of arithmetic flow of NeuDATool.
}
}
\end{center}
\end{figure}

\begin{enumerate}[I.]
\item Define atoms, molecules, and the simulation box with inheritance in \cpp. Three basic
\cpp\ classes have been designed to help users to define their special sub-classes or objects:

$\bullet$ Atom: It is used to define an atom type in a simulation. Some basic atomic properties,
such as the atom name, element name, isotope name, coordinate position, neutron scattering length,
are defined. In the class, the coordinate position of an atom is defined using the Hep3Vector
class of the class library for high energy physics (CLHEP)\cite{CLHEP_01,CLHEP_02} because the
Hep3Vector has abundant functions to perform transition, rotation, distance, and angle calculations.
Users need to initialize atoms in molecular objects rather than define new atom classes.

$\bullet$ Molecule: It is used to define molecules, intramolecular potentials and their movements,
such as translation, rotation, etc. Users need to define new molecular classes through
inheritance and can try special inter- and intramolecular movements for their analyses. This
class makes the program very flexible and user-friendly.

$\bullet$ SimBox: It is used for generating a simulation box. Users need to define a subclass for their
simulation. In the subclass, users only need to edit an initial function with defined molecules.
Users can use any suitable algorithm to place the molecules in the model box for generating an initial
conformation.

\item Recompile and Run Program.

\item MC: The program performs MC simulations with reference potentials until equilibrium is reached.
The program moves molecules or atoms in sequence or randomly. The potential energy variation of the
simulation box ($\Delta U = U_{after} - U_{before}$) is used as the movement acceptance criterion.
If $\Delta U < 0$, the movement is accepted. If $\Delta U > 0$, the movement is accepted with a
probability $e^{-\Delta U/kT}$.

\item  Calculate $g(r)$, $S(Q)$, and EMP:  When MC reaches equilibrium, the program calculates
the difference of the neutron structural factor $\Delta S(Q)$ between the simulation $S_{sim}(Q)$
and experiment $S_{exp}(Q)$. EMP is calculated by applying the Fourier transform to $\Delta S(Q)$.
The program adds the EMP and RP together as the updated potential to perform the EPMC simulation.

\item  EPMC: The empirical potential Monte Carlo (EPMC) algorithm is very similar to MC with the exception
that when the simulation reaches equilibrium, the program calculates the EMP again, adds it to the
previous potential, and  performs the simulation with the updated potential. When the EMP attains
a very low value, the program starts to accumulate simulation data.

\item  Accumulate $g(r)$ and $S(Q)$: The program still perform EPMC to accumulate simulation data to
improve statistics until smooth $g(r)$ and $S(Q)$ curves are obtained. With the exception of $g(r)$ and
$S(Q)$, the program outputs a coordinate file which includes all the atoms in a text format, as used
in the GROningen machine for chemical simulations (GROMACS) (with the suffix of .gro)\cite{GROMACS_01},
or as used in the large-scale atomic/molecular massively parallel simulator (LAMMPS) (with the suffix
of .xyz)\cite{LAMMPS_01}. Accordingly, this file can be input to GROMACS or LAMMPS to calculate enthalpy,
entropy, etc., and also can be visualized with visual molecular dynamics (VMD)\cite{VMD_01,VMD_02}.
Users can define, calculate, and output any interesting variables which are related directly with the
atomic structure of the sample by adding new output functions.
\end{enumerate}

\subsection{Acceleration method}

\begin{figure}
\begin{center}
\includegraphics[height=0.5\linewidth,width=0.85\linewidth]
{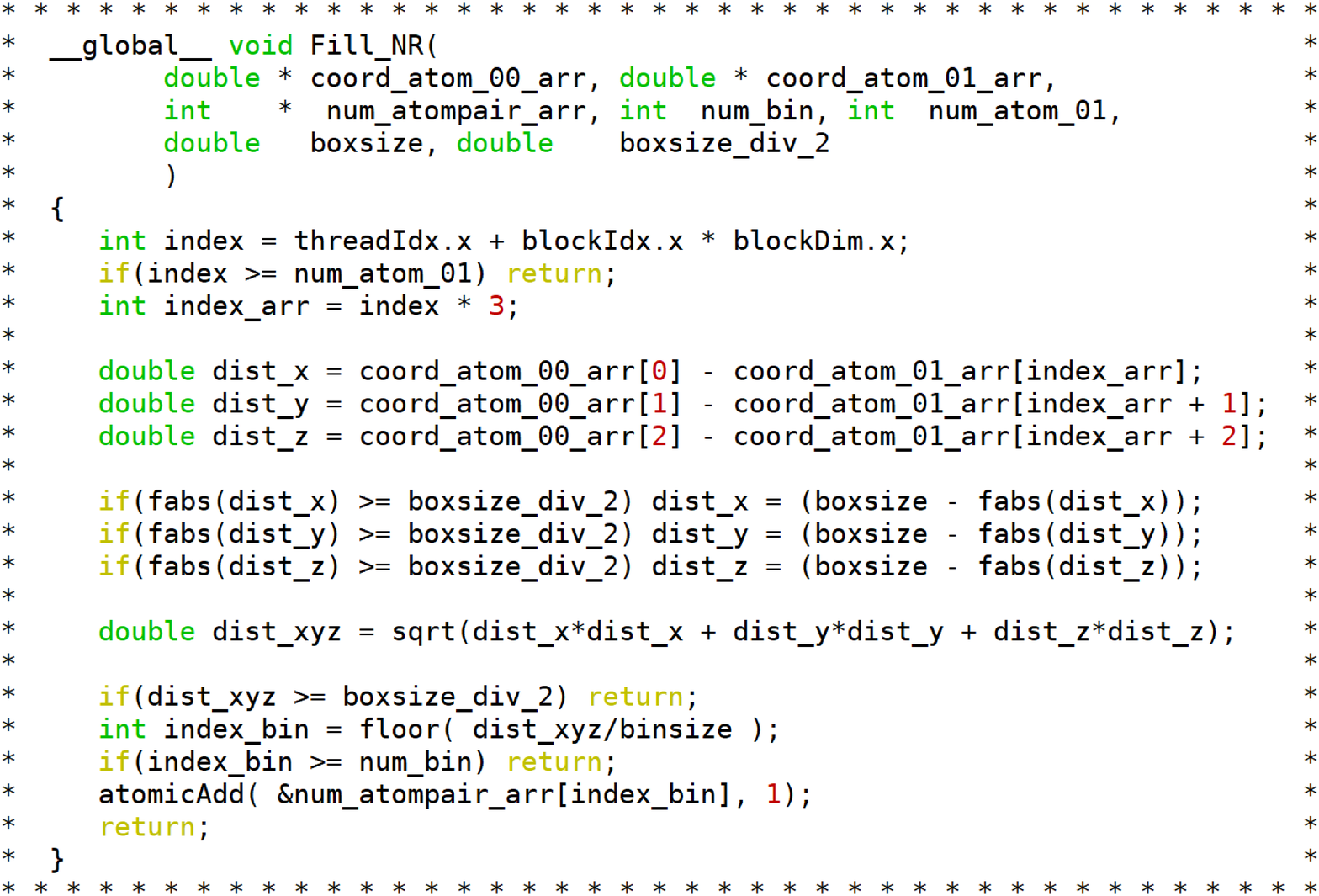}
\renewcommand{\figurename}{Fig.}
\textcolor[rgb]{0,0,1.0} {\figcaption
{
  \label{fig:figure_gpu_code} The CUDA C  code for calculate $N(r)$.
}
}
\end{center}
\end{figure}

The computation consumptions of $\Delta N(r)$ and $N(r)$ are in approximate proportion to the atom
number in the simulation box and its second order, respectively. In general, the amount of atoms in a
simulation box is larger than ten thousand. $\Delta N(r)$ needs to be recalculated
after every MC simulation step for $\Delta U$ calculation, while $N(r)$ needs to be
recalculated after every MC/EPMC equilibrium for $g(r)$ and $S(Q)$ update. Thus, these algorithms
represent the highest consumption of the program's calculation capacity.
A graphics processing unit (GPU) has thousands parallel threads, so it is a suitable candidate to
accelerate the calculation of $\Delta N(r)$ and $N(r)$. Fig.\ref{fig:figure_gpu_code} shows the
CUDA C kernel function used to calculate $N(r)$ invoking GPU acceleration.

For the implementation of across nodes in a parallel configuration, mpich2 is
used in the program\cite{MPICH_01}. Mpich2 is based on  the MPI\cite{MPI_01}
standard and supports point-to-point and collective data communication among different
nodes. Thus, it is highly efficiency in this program.
For implementing a shared memory, multithread, parallel configuration within a
computer or server node, Open Multi-Processing  (OpenMP) is used in the \cpp{ } program\cite{OpenMP_01}.
The OpenMP syntax supports the setting of a thread number dynamically, and a thread
number cannot be known in advance in most cases. Thus, it is very convenient in programming.
NeuDATool uses OpenMP and MPI API to distribute different $\Delta N(r)$/$N(r)$ to different GPU
cards belong to a computer node or different nodes in a computer cluster. In addition,
CPU serial algorithm version is edited for the programs wide applicability.

\section{Performance Test}

Neutron scattering data of full hydrogen, full deuterated, half deuterated H$_2$O and
amorphous silica (SiO$_2$) samples at ambient temperature and pressure are used to test the
performance of NeuDATool regarding the correctness and computational speed.

\subsection{Correctness}

Experimental neutron data of full hydrogen  water, full deuterated water,
half deuterated water, and SiO$_2$ distributed with EPSR\cite{EPSR_Web}
and GudRun\cite{GUDRUN_Web,GUDRUN_01} in the ISIS website are used to
test the program.

For simulating the sample of water, we define a class to describe molecule of H$_2$O
by inheriting form molecule basic class. The molecule structure is maintained with
harmonic oscillator potentials between every two atoms. Three different random movements
\ie, H$_2$O's translation, H$_2$O's rotation, and atom's (H or O) translation, are
implement in the simulation. The EMP ($U^{EMP}$) of the atom type pairs of the sample
are calculated with \textbf{Eq.\ref{eq:emp_cal}}.\cite{EPSR_03}
{\wuhao
\begin{equation}
\label{eq:emp_cal}
U^{EMP}_{j}(r) = \frac{1}{4\pi\rho} \sum_{i= 1,M}
 w_{ji}^{-1}\int_{0}^{\infty} \Delta S_{i}(Q) e^{-iQr}dQ
\end{equation}
}
Where $U^{EMP}_{j}(r)$ is the EMP of different the atom pairs, \ie, O-O, O-H, and H-H in water;
$\Delta S_{i}(Q)$ is the different neutron structure factor between experiment and simulation of
full hydrogen, full deuterated, and half deuterated H$_2$O; $w_{ji}^{-1}$ is the inverse
of weighted factor matrix in \textbf{Eq.(\ref{eq:cs_cal_liq})}. In water samples, the number of
neutron scattering profiles and atom type pairs are the same and equal to 3, So we employed matrix
Invert() function provided by CLHEP\cite{CLHEP_01,CLHEP_02} to calculate the inverse matrix.

In the simulation of SiO$_2$, we define two classes to describe Si and O as single atom molecules.
The most difference from water is that there are only one neutron scattering profile but 3 different
atom type pairs, \ie, Si-Si, Si-O, and O-O. As do in EPSR\cite{EPSR_03}, we use a $3\times3$
unit matrix to enlarge the $(1\times 3)$ weighted factor matrix $w_{ij}$ to a $(4\times 3)$
matrix $w_{ij}'$, then use a Monte-Carlo method to calculate the pseudo-inverse matrix $w'^{-1}$.

\begin{figure}[h!]
\centering
\includegraphics[width=0.48\linewidth, height=0.32\linewidth]{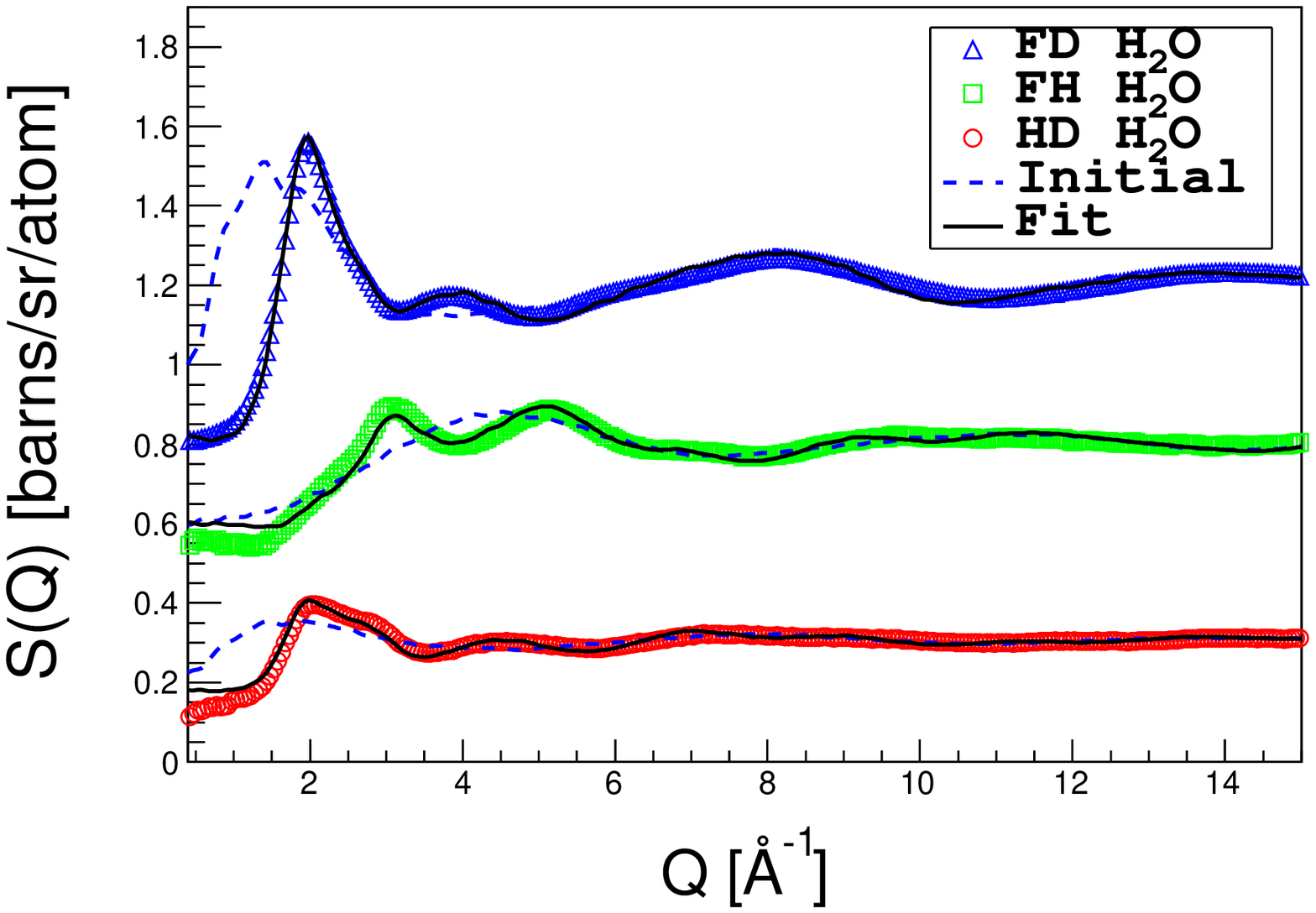}
\includegraphics[width=0.48\linewidth, height=0.32\linewidth]{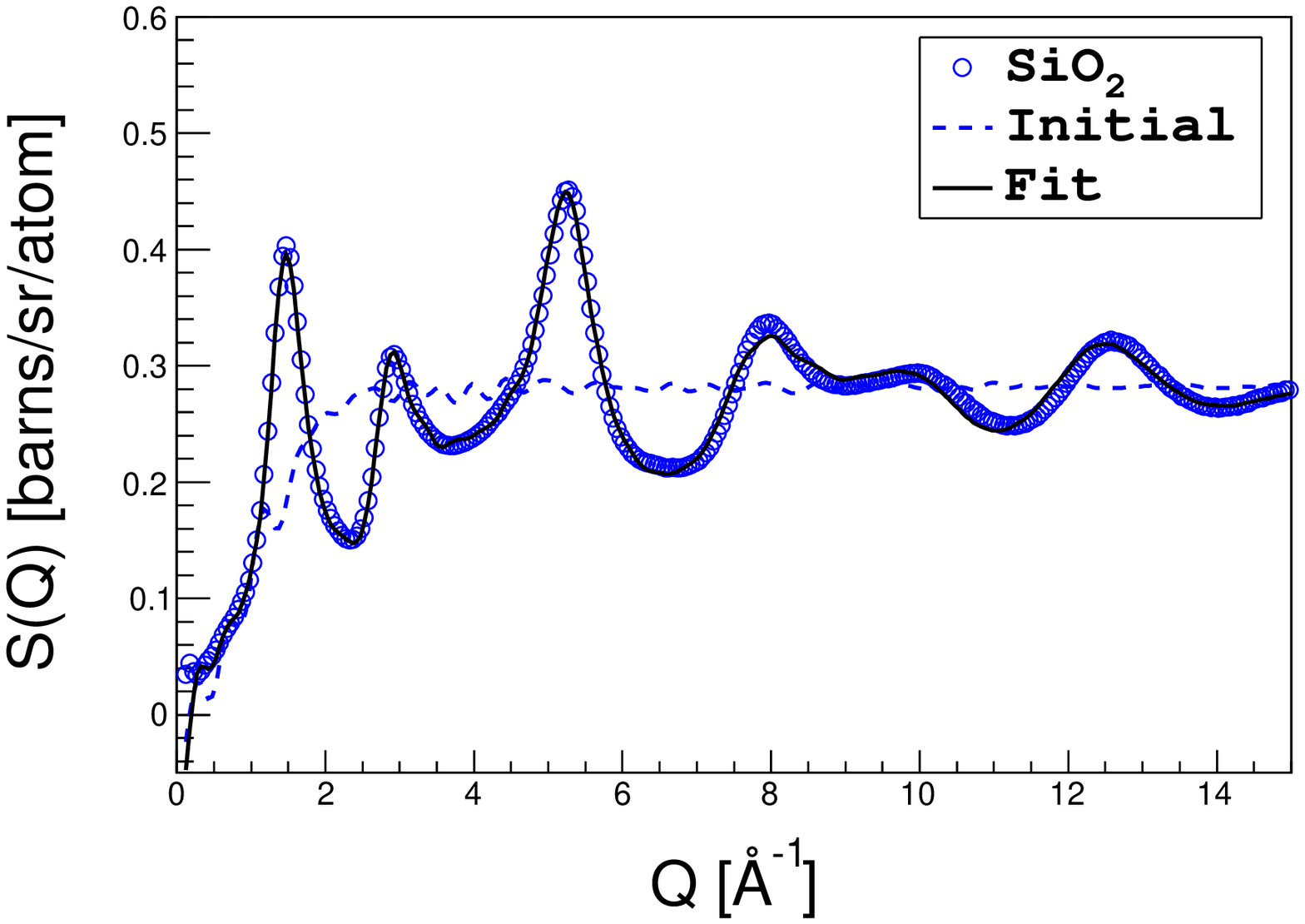}
\textcolor[rgb]{0,0,1.0} {%
\caption{\label{fig:nsf_ndatool_fit}
Neutron scattering spectral comparison between NeuDATool and the experimental sample. Left:
H$_2$O samples.  Right: Amorphous SiO$_2$ sample.
The points denote the experiment data, while the solid lines denote the NeuDATool simulations.
The dashed lines denote the random initial simulation boxes.
 }}
\end{figure}

Fig.\ref{fig:nsf_ndatool_fit} shows the NeuDATool simulation results. They are consist
with the experiments. It confirms that the distributions of small molecules observed in neutron
scattering experiments have been reasonably represented by NeuDATool simulation.

\begin{figure}[h!]
\centering
\includegraphics[width=0.8\linewidth, height=0.5\linewidth]{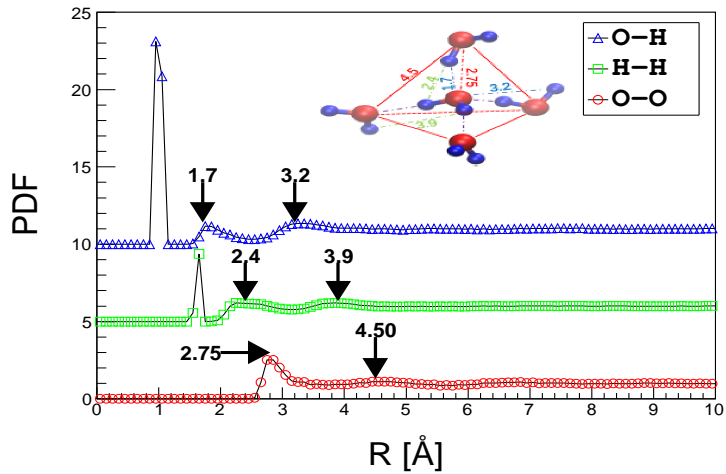}
\textcolor[rgb]{0,0,1.0} {%
\caption{\label{fig:pdf_h2o} PDF distributions of O-H, H-H, and O-O from  NeuDATool simulation for water,
and the inset is a cartoon of the liquid water structure\cite{water_03,water_01,water_02}.
}}
\end{figure}

To further verify the reliability of the simulation, we compared the PDF distribution
of H$_2$O from NeuDATool with other studies. Fig.{\ref{fig:pdf_h2o}} shows the PDF
distribution from NeuDATool simulation. Liquid water is a tetrahedrally random network.
On average, 3.5 water molecules form hydrogen bonding with a center one.
The results are consist with the studies
by Kusalik, Head-Gordona and Soper\cite{water_03,water_01,water_02}.

All of those prove NeuDAToool can reconstruct the atomic structures of experimental
samples correctly based on the neutron diffraction profile.

\begin{figure}[h!]
\centering
\includegraphics[width=0.8\linewidth, height=0.4\linewidth] {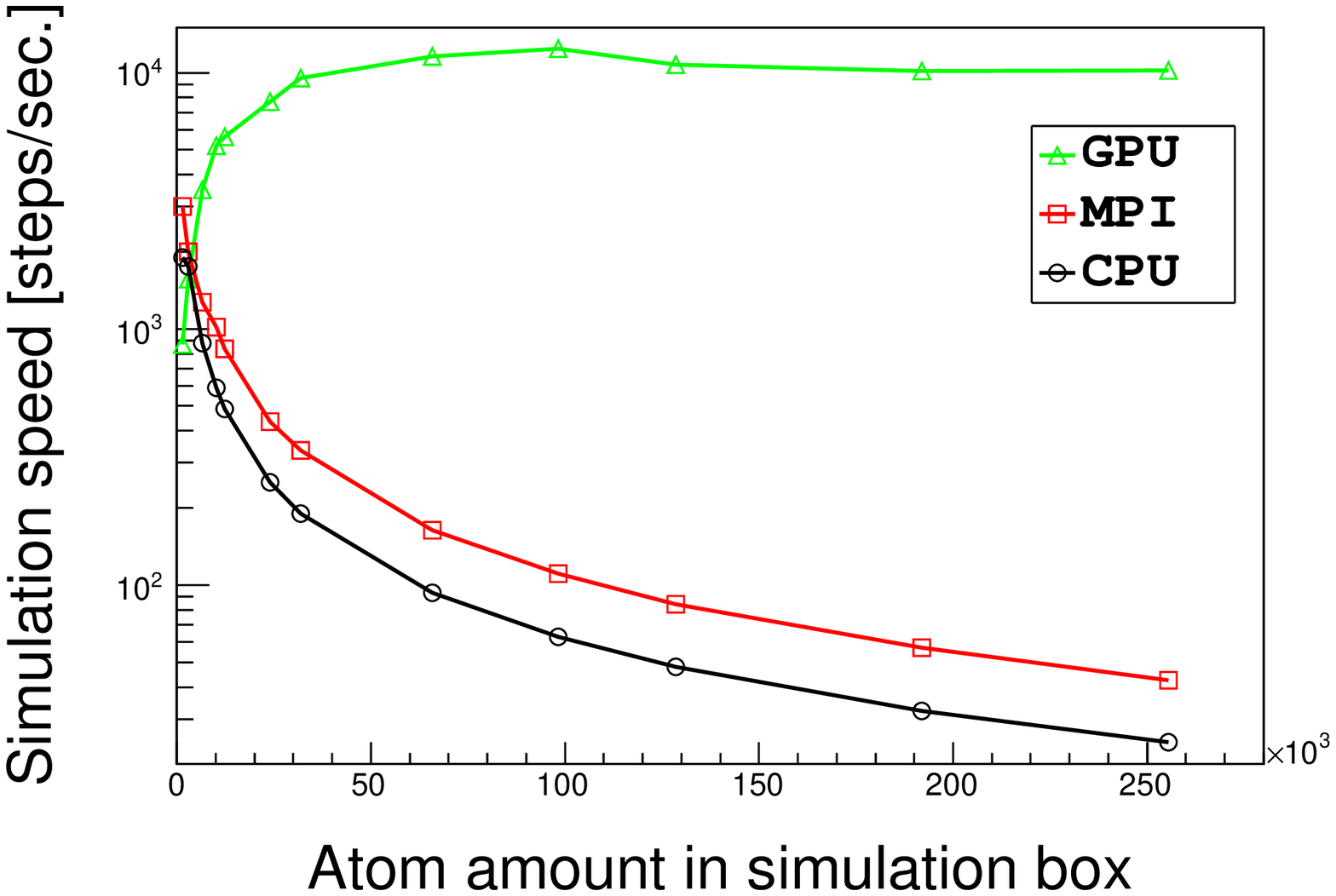}
\textcolor[rgb]{0,0,1.0} {%
\caption{\label{fig:cpu_gpu_speed} Calculation speed comparison among serial algorithm, MPI parallel and GPU acceleration algorithms. The X coordinate denotes the atomic number of the simulation boxes, while the Y coordinate denotes the simulation steps in units of seconds.
}}
\end{figure}

\subsection{Computational Speed}

A small computer cluster
is used to test the speed performance of different methods. The cluster uses
CentOS 7.3 as the operating system and has two nodes. Each node has two Intel
Xeon Scalable Gold 6126 CPU (two Skylake--SP architectures, 12 cores, 24 threads，
2.6 GHz, Turbo 3.7 GHz，and a 19.25 MB L3 Intel smart cache), two Nvidia Tesla V100
calculation GPU card and 128 GB double data rate (DDR4) error correcting code (ECC)
registered shared memory. The two nodes are connected with an InfiniBand (IB)
connector (data transmission speed can increase up to 56 Gb/s).

A speed comparison of these acceleration methods is shown in Fig.\ref{fig:cpu_gpu_speed},
and a more detailed quantitative comparison is listed in Tab.\ref{tab:tab_speed}.
The water experiment data used in the speed test is the same as in the correctness test.
As shown in the figure and table, mpich2  can improve the speed based on the ratio of
nodes or thread numbers, while the GPU can provide an excellent acceleration ratio.
Most importantly, with the GPU acceleration, the program can simulate a system comprising
$>$ 1 million atoms. This is an essential improvement because it allows the program to simulate
systems larger than 200 \AA, so that it can analyze samples with macromolecules in
all atomic models in the future.

\begin{table}
\begin{center}
\tabcaption{
\label{tab:tab_speed}
Simulation speeds with GPU acceleration, MPI parallel and CPU serial algorithms
[steps/sec.].
  }
{\liuhao
\begin{tabular}{
c| C{1.6eM}  C{1.6eM}   C{1.6eM}   C{1.6eM}   C{1.6eM} C{1.6eM}  C{1.6eM}   C{1.6eM}}
\toprule
\multicolumn{1}{c|}{Atomic number } &
\multicolumn{1}{c}{$3\times10^3$ } &
\multicolumn{1}{c}{$3\times10^4$ } &
\multicolumn{1}{c}{$1\times10^5$ } &
\multicolumn{1}{c}{$2.5\times10^5$ } &
\multicolumn{1}{c}{$1.2\times10^6$ } &
\multicolumn{1}{c}{$3\times10^6$ } &
\multicolumn{1}{c}{$1\times10^7$ } &
\multicolumn{1}{c}{$3\times10^7$ }
\\
\hline
 GPU & 1562  & 9507  & 12412   & 10189 & 3917  & 1963  & 757.0 & 340.3
\\
 MPI & 2000  & 334.9  & 110.5  & 42.57 & ---  & ---  & --- & ---
\\
 CPU & 1754  & 189.8  & 62.67  & 24.30 & ---  & ---  & --- & ---
\\
\bottomrule
\end{tabular}
}
\end{center}
\end{table}

\section{Conclusions}

The neutron scattering data analysis software NeuDATool is programmed with the
object-oriented language \cpp. It  makes  the program  flexible and friendly
to users who need to define special molecules and MC random movement patterns.
Potential functions and the corresponding parameters of the atomic force field
can be modified or added by editing the \cpp { } head and source files.
In addition, \cpp{ } is an easy-to-read, high-level computer language, so users
can try new algorithms and program flows to improve their analyses, and to
calculate and output any important variables.

With the exception of parallel nodes within the server with OpenMP, parallel
cross-different nodes of a computer cluster, and GPU hardware acceleration are
supported. Specifically, with GPU acceleration,
the calculation speed is improved considerably, so the program has the
capacity to analyze disordered macromolecular samples and nanoparticles of all
atomic models in the future.

Although the program is flexible for users and has a powerful calculation
capacity, it was tested with a very limited number of control samples. Accordingly,
in its current form, it is not a fully functional software package. The authors
aspire to release it as an open-source toolkit framework for public use by
interested scientists. In this sense, users will be able to contribute numerous
new molecular classes, algorithms, and analyses routines in the future to make
the program more powerful.

\section{Acknowledgements}
This work was supported by the National Key Research and Development Program
of China [grant number 2017YFA-0403703]; and the National Natural Science
Foundation of China (NSFC) [grant number U1830205, 21674020]. The authors thank
the staff of the ISIS Disordered Materials Group for the explanations of the
principles of EPSR and for providing experimental neutron diffraction data
for testing of the program.

\rule{\textwidth}{0.2mm}

\bibliographystyle{elsarticle-num}
\bibliography{NeuDATool_bib}


\end{document}